\documentstyle[preprint,aps]{revtex}
\def\etwa{\mbox{${}_{\sim}$}}
\def\groesser{\mbox{${}^{>}$}}
\def\kleiner{\mbox{${}^{<}$}}
\newlength{\blib}
\def\gesim{\mbox{$\,\groesser\hspace{-\blib}\etwa\,$}}
\def\lesim{\mbox{$\,\kleiner\hspace{-\blib}\etwa\,$}}
\begin{document}
\draft
\title{Electron--phonon interaction and optical spectra of metals}
\author{H.\ J.\ Kaufmann\thanks{corresponding author; Tel. 
+44-1223-333409, Fax +44-1223-333450, e-mail hjk22@cus.cam.ac.uk}}
\address{Interdisciplinary Research Centre in Superconductivity,
University of Cambridge, Madingley Road, Cambridge CB3 0HE, UK and
Department of Earth Sciences, University of Cambridge,
Downing Street, Cambridge CB2 3EQ, UK.}
\author{E.\ G.\ Maksimov}
\address{P. N. Lebedev Physical Institute, 117924 Moscow, Russia.}
\author{E.\ K.\ H.\ Salje}
\address{Interdisciplinary Research Centre in Superconductivity,
University of Cambridge, Madingley Road, Cambridge CB3 0HE, UK and
Department of Earth Sciences, University of Cambridge,
Downing Street, Cambridge CB2 3EQ, UK.}
\begin{tighten}
\maketitle
\begin{abstract}
Observed optical reflectivity in the infrared spectral region is compared
with theoretical predictions in a strongly coupled electron--phonon
system. Starting from a Fr\"ohlich Hamiltonian, the spectral functions
and their temperature dependence are derived. A full analysis including
vertex corrections leads to an expression for the optical conductivity 
$\sigma(\omega)$ which can be formulated in terms of
the well known optical conductivity for a quasi--isotropic
system without vertex corrections. A numerical comparison between the
full result and the so--called ``extended'' Drude formula, its 
weak coupling expansion, show little difference over a wide range of 
coupling constants. 
Normal state optical spectra for the high-${\rm T_c}$ superconductors 
YBa$_2$Cu$_3$O$_7$ and La$_{2-x}$Sr$_{x}$CuO$_4$ at optimal doping
are compared with the results of model calculations. 
Taking the plasma frequency and
$\epsilon_{\infty}$ from band structure calculations, the model
has only one free parameter, the electron--phonon coupling constant
$\lambda$.  
In both materials the overall behaviour of the reflectivity can be 
well accounted for over a wide frequency range. Systematic 
differences exist only in the mid--infrared region. They become more
pronounced with increasing frequency, which indicates that a detailed
model for the optical response should include temperature dependent
mid--infrared bands.
\end{abstract}
\pacs{Keywords: optical properties, electron--phonon interaction,
high--${\rm T_c}$ superconductors}
\section{Introduction}
The optical spectra of metals in the infrared (IR) spectral region depend
sensitively on the interaction between electrons and phonons.
Deviations from the theoretical spectrum without any phonon contribution
stem from the so--called Holstein mechanism \cite{holstein} in which the
incident photon is absorbed in a second--order
process involving creation of both a phonon and an electron--hole pair.
The detailed description of this mechanism was given by Allen 
\cite{allen}. Despite the fact that the physical mechanism of
electron--phonon coupling is rather well understood for almost three decades,
little process was made in the systematic experimental study of
coupling effects in optical (IR) spectra \cite{motulevich}, nor has the
phenomenology been explored beyond the lowest order effects.
The lack of systematic experimental investigations appears to be related to the 
following two reasons:

Firstly, historically measurements of optical spectra of metals where 
limited to photon energies $\omega \gesim 0.05 $ eV \cite{motulevich}. 
We shall argue in this paper that the effects of the 
electron--phonon interaction on the optical spectra of ordinary metals are rather 
weak at these energies.  The optical conductivity in this spectral range
can usually be described in the framework of the Drude formula
\begin{equation}\label{drude}
\sigma(\omega)=\frac{\omega_{\rm pl}^{2}}{4\pi}\,\frac{1}{-{\rm i}\/\omega+1/\tau}
\end{equation}
where $1/\tau$ is the relaxation rate of electrons due to their interaction
with impurities and phonons. It can be calculated using the commonly
adapted 
Bloch--Gr\"uneisen--type formula.

Secondly, the measurements of the optical conductivity are complicated in 
ordinary metals by the anomalous skin effect which is mostly present at low
temperatures and in the far--infrared (FIR) spectral region. It leads to serious 
difficulties in the interpretation of reflectivity and absorbtivity 
measurements. Only a few observations \cite{joyce,bednorz} are known to us where Holstein
processes where identified in the normal state of ordinary metals.

The discovery of the high-T$_{\rm c}$ superconductors (HTSC) has dramatically
changed the experimental situation. First of all, the experimental methods were 
improved radically by extending the accessible energy range down to 
$\approx 10$ cm$^{-1}$ and by increasing the accuracy of the measurements. 
Secondly, it appears that HTSC systems allow the observation of
the electron--phonon interaction in the optical spectra more easily and
more clearly than it was possible in the ordinary metals. Thirdly, there
is no anomalous skin effect in these systems for light with electric field 
parallel to the Cu--O planes.

It is the purpose of this paper to review the theoretical situation.
We then extend the treatment of the Fr\"ohlich Hamiltonian to
include vortex corrections.
Theoretical predictions are then compared with experimental observations following earlier
attempts to connect the IR reflectivity and absorption spectra of
HTSCs with features of strongly interacting electron--phonon systems
\cite{shulga,oleg,holger}.

We shall show that most but not all observations can be described rather well 
within such a scheme and identify some pertinent open questions.

\section{Derivation of the optical conductivity in the framework of
the Fr\"ohlich Hamiltonian}\label{formulae}
We start from the description of metals with electron--phonon interaction by
the standard Fr\"ohlich Hamiltonian
\begin{eqnarray}\label{froehlich}
H &=& \sum\limits_{{\bf k},i} \epsilon_{{\bf k},i}\,
c^{\dagger}_{{\bf k},i}\,c_{{\bf k},i} +
\sum\limits_{{\bf k},{\bf q},i,i^{\prime},\lambda}
g_{\bf k}({\bf q},i,i^{\prime},\lambda)\,c^{\dagger}_{{\bf k},i}\,
c_{{\bf k}+{\bf q},i^{\prime}} \times
\\\nonumber
&&\times
\left( b^{\dagger}_{{\bf q}\lambda} + b_{-{\bf q}\lambda}\right) +
\sum\limits_{{\bf q},\lambda} \omega_{{\bf q}\lambda}
b^{\dagger}_{{\bf q}\lambda}\,b_{{\bf q}\lambda}\;. 
\end{eqnarray}
Here the first term is the kinetic energy of an electron with given momentum
${\bf k}$ and band index $i\/$, the last term is the energy of the phonon 
with momentum ${\bf q}$ and mode $\lambda$.
The second term represents the electron--phonon interaction, where
$g_{\bf k}({\bf q},i,i^{\prime},\lambda)$ is the matrix element of this interaction.
The use of this Hamiltonian for the self--consistent calculation of the 
electron and phonon Green's functions cannot be rigourously justified in the general
case. It was shown \cite{allen1,zhenya,rainer} that this Hamiltonian can 
be used for the calculation of the influence of the electron--phonon interaction 
on the electronic properties for systems where no low--energy collective excitations 
of electrons are present. First--principle calculations
\cite{zhenya} of the physical properties of a number of ordinary
metals have shown that the Fr\"ohlich Hamiltonian is a very good
starting point for the analysis of all features related to
electron--phonon interactions.
In HTSC materials we can not expect this Hamiltonian to describe the optical
response completely because additional low
energy excitations (e.\,g.\ spin and/or charge excitations) occur.
However,
these excitations contribute little to the general trend of 
the ``anomalous'' optical properties of HTSC systems at high energies and
are simply ignored in this study.
The question we wish to answer here is the following: to what extend can 
observed optical spectra in the normal state be understood in terms
of the most simple model of electron--phonon interaction as represented
by equation (\ref{froehlich}).

The many--body electron--phonon interaction are usually calculated by Green's 
function method \cite{abrikosov}. 
Let us introduce the electron and phonon one--particle thermodynamic Green's 
functions
\begin{equation}
G_{i}({\bf k},\tau)= -\langle {\rm T}_{\tau}\,
c^{\dagger}_{{\bf k},i}(\tau)\,c_{{\bf k},i}(0)
\rangle\;\,
\end{equation}
and
\begin{equation}
D_{\lambda}({\bf q},\tau)= -\langle {\rm T}_{\tau}\,
b^{\dagger}_{{\bf q}\lambda}(\tau)\,b_{{\bf q}\lambda}(0)
\rangle\;.
\end{equation}
The Wick operator ${\rm T}_{\tau}$ reorders the operators following it
in such a way that $\tau$ increases from right to left. For non--interacting 
particles the Fourier components of the Green's functions, 
$G_{i}({\bf k},{\rm i}\/\omega_n)$ and $D_{\lambda}({\bf q},{\rm i}\/\omega_\nu)$, 
have the very simple form
\begin{equation}\label{freeelectron}
G_{i}^{0}({\bf k},{\rm i}\/\omega_n) = \frac{1}{{\rm i}\/\omega_n -
\epsilon_{{\bf k}, i}}
\end{equation}
and
\begin{equation}
D_{\lambda}^{0}({\bf q},{\rm i}\/\omega_\nu) = 
\left(\frac{1}{{\rm i}\/\omega_\nu - \omega_{{\bf q}\lambda}}
- \frac{1}{{\rm i}\/\omega_\nu + \omega_{{\bf q}\lambda}}\right)\;.
\end{equation}
Here ${\rm i}\/\omega_n = (2n+1)\pi{\rm T}$ 
and ${\rm i}\/\omega_\nu = 2\nu\pi{\rm T}$
are the Matsubara frequencies for fermions
and bosons, respectively, and ${\rm T}$ is the
temperature.
The value
$\epsilon_{{\bf k}, i}$ is the electron band energy and, correspondingly, 
$\omega_{{\bf q}\lambda}$ is the phonon energy for the $\lambda$th mode.
In the following we do not consider the renormalisation of the phonon 
Green's function due to the electron--phonon interaction. For
convenience we present the phonon function 
$D_{\lambda}({\bf q},{\rm i}\/\omega_\nu)$ in the spectral form
\begin{equation}
D_{\lambda}({\bf q},{\rm i}\/\omega_\nu) = 
\frac{1}{\pi}\,\int\limits_{0}^{\infty}{\rm d}\Omega\: {\rm Im}\,
D_{\lambda}({\bf q},\Omega)
\left(\frac{1}{{\rm i}\/\omega_\nu - \Omega}
- \frac{1}{{\rm i}\/\omega + \Omega}\right)\;.
\end{equation}
For the non--interacting case the spectral density 
${\rm Im}\,D_{\lambda}({\bf q},\Omega)$ has the form
\begin{equation}
{\rm Im}\,D^{0}_{\lambda}({\bf q},\Omega) = \pi\,\delta(\Omega - 
\omega_{{\bf q}\lambda})\;.
\end{equation}
In the following we tread the spectral function
${\rm Im}\,D_{\lambda}({\bf q},\Omega)$ as an experimental quantity and
calculate the influence of the 
electron--phonon interaction on the electronic properties. 
These properties are reflected in
the one--particle electron Green's function
and the optical conductivity $\sigma_{\alpha\beta}(\omega)$.
The one--particle Green's function for electrons in
the presence of the electron--phonon interaction 
\cite{migdal}
can be written as
\begin{equation}
G^{-1}({\bf k},{\rm i}\/\omega_n) = G^{-1}_0({\bf k},{\rm i}\/\omega_n) - 
\Sigma({\bf k},{\rm i}\/\omega_n)
\end{equation}
where $\Sigma({\bf k},{\rm i}\/\omega_n)$ is the electron self--energy part.
One of the main results of Migdal \cite{migdal} is that the electron self--energy
can be calculated using the simplest first--order approximation
in the electron--phonon interaction, neglecting all vertex corrections
as being small of the order of $\omega_{\rm D}/\epsilon_{\rm F}$.
Here $\omega_{\rm D}$ is a characteristic phonon energy and $\epsilon_{\rm F}$
is the Fermi energy of the electrons. An analytical expression of 
$\Sigma({\bf k},{\rm i}\/\omega_n)$ is
\begin{eqnarray}
\Sigma_i({\bf k},{\rm i}\/\omega_n) &=& -{\rm T}\,
\sum\limits_{\omega_\nu}
\sum\limits_{{\bf k}^{\prime},i^{\prime},\lambda}
\left| g_{{\bf k}}({\bf k} - {\bf k}^{\prime},
i, i^{\prime}, \lambda)\right|^2
D_{\lambda}({\bf k} - {\bf k}^{\prime},{\rm i}\/\omega_\nu)\,\times
\\\nonumber&&\times
G({\bf k}^{\prime},{\rm i}\/\omega_n-{\rm i}\/\omega_\nu)\;.
\end{eqnarray}
The summation over the momentum ${\bf k}^{\prime}$ is 
represented in integral form
\begin{equation}
\sum\limits_{{\bf k},i} = \sum\limits_{i}\int\limits_{-\infty}^{\infty}
{\rm d}\/\epsilon\: \sum\limits_{\bf k}\delta(\epsilon - \epsilon_{{\bf k},i})
\end{equation}
so that
$\Sigma_i({\bf k},{\rm i}\/\omega_n)$ becomes
\begin{eqnarray}\label{phonongf}
\Sigma_i({\bf k},{\rm i}\/\omega_n) &=& -{\rm T}\,
\sum\limits_{\omega_\nu}\,
\int\limits_{-\infty}^{\infty}{\rm d}\/\epsilon\: 
\sum\limits_{{\bf k}^{\prime},i^{\prime},\lambda}
\left| g_{{\bf k}}({\bf k} - {\bf k}^{\prime},
i, i^{\prime}, \lambda)\right|^2
\delta(\epsilon - \epsilon_{{\bf k}^{\prime},i^{\prime}})\times
\\\nonumber &&\times
\int\limits_{0}^{\infty}{\rm d}\/\Omega\:
{\rm Im}\,D_{\lambda}({\bf k}-{\bf k}^{\prime},\Omega)\, 
G({\bf k}^{\prime},{\rm i}\/\omega_n-{\rm i}\/\omega_\nu)
\times\\\nonumber &&\times
\left(\frac{1}{{\rm i}\/\omega_\nu - \Omega}
- \frac{1}{{\rm i}\/\omega_\nu + \Omega}\right)\;.
\end{eqnarray}
The analysis of (\ref{phonongf}), as performed by Migdal,
shows that the essential values of ${\rm i}\/\omega_n$ 
and ${\rm i}\/\omega_\nu$ are of order $\omega_{\rm D}$.
This means that small frequencies of order $\omega_{\rm D}$ are
also significant for $\epsilon$. In this case $\epsilon$ can
be neglected in $\delta(\epsilon - \epsilon_{{\bf k},i})$.
The electron Green's function is used in the form given by Eq.\ 
(\ref{freeelectron}) leading to
\begin{eqnarray}\label{phonongf2}
\Sigma_i({\bf k},{\rm i}\/\omega_n) &=& -{\rm T}\,
\sum\limits_{\omega_\nu}
\sum\limits_{{\bf k}^{\prime}}
\delta(\epsilon_{{\bf k}^{\prime}})
\,\int\limits_{0}^{\infty}{\rm d}\/\Omega\:
\alpha^{2}_{i}({\bf k},{\bf k}^{\prime},\Omega)F(\Omega)
\left(\frac{1}{{\rm i}\/\omega_\nu - \Omega}
- \frac{1}{{\rm i}\/\omega_\nu + \Omega}\right)
\times\\\nonumber &&\times
\int\limits_{-\infty}^{\infty}{\rm d}\/\epsilon\:
\frac{1}{{\rm i}\/\omega_n -{\rm i}\/\omega_\nu -\epsilon}\;,
\end{eqnarray}
where the spectral function of the electron--phonon interaction
\begin{eqnarray}
\alpha^{2}_{i}({\bf k},{\bf k}^{\prime},\Omega)F(\Omega)
&=&
\sum\limits_{i^{\prime},\lambda}
\left| g_{{\bf k}}({\bf k} - {\bf k}^{\prime},
i, i^{\prime}, \lambda)\right|^2 
{\rm Im}\,D_{\lambda}({\bf k}-{\bf k}^{\prime},\Omega)
\end{eqnarray}
was introduced. 
The sum over the $\omega_\nu = 2\pi\nu\/{\rm T}$
can be easily performed with the result
\begin{eqnarray}\label{sumresult}
-{\rm T}\,\sum\limits_{\omega_\nu}
\left(\frac{1}{{\rm i}\/\omega_\nu - \Omega}
- \frac{1}{{\rm i}\/\omega_\nu + \Omega}\right)
\frac{1}{{\rm i}\/\omega_n -{\rm i}\/\omega_\nu -\epsilon}
&=&\hspace*{2cm}
\end{eqnarray}
\begin{eqnarray}\nonumber \hspace*{2cm}&=&
\frac{N(\Omega)+1-f(\epsilon)}{{\rm i}\/\omega_n-\Omega-\epsilon}
+
\frac{N(\Omega)+f(\epsilon)}{{\rm i}\/\omega_n+\Omega-\epsilon}
\;,
\end{eqnarray}
where $N(\Omega)$ and $f(\epsilon)$ are the Bose and Fermi function,
respectively. To obtain the one--particle Green's function describing the
electron excitation spectrum we use the analytic continuation
of the self--energy on the imaginary axis $\omega$.
This can easily be done by changing ${\rm i}\/\omega_n$ in 
Eq.\ (\ref{sumresult}) to $\omega$. Consequently, the final 
expression for the self--energy reads
\begin{eqnarray}\label{selfenergy}
\Sigma_i^{{\rm R},{\rm A}}({\bf k},\omega) &=&
\int\limits_{0}^{\infty}{\rm d}\/\Omega\:
\sum\limits_{{\bf k}^{\prime}}
\delta(\epsilon_{{\bf k}^{\prime}})
\,\alpha^{2}_{i}({\bf k},{\bf k}^{\prime},\Omega)F(\Omega)
L(\omega\pm {\rm i}\/\delta,\Omega)
\;,
\end{eqnarray}
where 
\begin{eqnarray}\label{l}
L(\omega\pm {\rm i}\/\delta,\Omega)
&=&
\int\limits_{-\infty}^{\infty}{\rm d}\/\epsilon\:
\left[
\frac{N(\Omega)+1-f(\epsilon)}{\omega-\Omega-\epsilon\pm {\rm i}\/\delta}
+
\frac{N(\Omega)+f(\epsilon)}{\omega+\Omega-\epsilon\pm {\rm i}\/\delta}
\right]
\;.
\end{eqnarray}
The function $\Sigma^{{\rm R}\,({\rm A})}({\bf k},\omega)$ 
denotes the retarded (advanced) self--energy which is an analytical 
function of the variable $\omega$ in the upper (lower) 
half of the complex plane.
The integral in (\ref{l}) can be evaluated analytically, yielding 
\begin{eqnarray}\label{l2}
L(\omega,\Omega)
&=&
-2\pi\/{\rm i}\/\left[ N(\Omega)+\frac12 \right] +
\Psi\left( \frac12 + {\rm i}\/\,\frac{\Omega-\omega}{2\pi\/{\rm T}}\right)
- 
\Psi\left( \frac12 - {\rm i}\/\,\frac{\Omega+\omega}{2\pi\/{\rm T}}\right)
\,,
\end{eqnarray}
where $\Psi(z)$ is the digamma function.
The self--energy $\Sigma_i^{{\rm R},{\rm A}}({\bf k},\omega)$ expressed by 
Eq.\ (\ref{selfenergy}) depends only on the direction of the momentum
${\bf k}$ on the Fermi surface. It is convenient to present
this dependence by expanding all functions involved 
in this expression over the complete and orthonormal set of functions
introduced by Allen \cite{allen2}. These so--called ``Fermi surface harmonics'',
$\Psi_j({\bf k})$, satisfy the condition
\begin{eqnarray}\label{fortho1}
\sum\limits_{\bf k}\Psi_j({\bf k})\Psi_{j^{\prime}}({\bf k})\,
\delta(\epsilon_{\bf k} - \epsilon ) &=& \delta_{jj^{\prime}}
N(\epsilon)\;,
\end{eqnarray}
where 
\begin{eqnarray}\label{fortho2}
N(\epsilon) &=& \sum\limits_{\bf k}\delta(\epsilon_{\bf k} - \epsilon ) 
\;.
\end{eqnarray}
In terms of this set we write
\begin{eqnarray}\label{sexpansion}
\Sigma_i^{{\rm R},{\rm A}}({\bf k},\omega)
&=&
\sum\limits_{j}\Sigma_{i,j}^{{\rm R},{\rm A}}(\omega)\Psi_j({\bf k})
\\\label{gexpansion}
G^{{\rm R},{\rm A}}_i({\bf k},\omega)
&=&
\sum\limits_{j}G_{i,j}^{{\rm R},{\rm A}}(\epsilon_{\bf k},\omega)
\Psi_j({\bf k})
\\\label{eliashbergfct}
\alpha^{2}_{j,j^{\prime}}({\bf k},\Omega)F(\Omega)
&=&
\sum\limits_{j,j^{\prime}}\sum\limits_{{\bf k}^{\prime},\lambda}
\delta(\epsilon_{{\bf k}^{\prime}})
\left\{
\left| g_{{\bf k}^{\prime}}({\bf k} - {\bf k}^{\prime},
i, i^{\prime}, \lambda)\right|^2 
{\rm Im}\,D_{\lambda}({\bf k} - {\bf k}^{\prime},\omega)
\right\}_{jj^{\prime}}
\times\\\nonumber &&\times
\Psi_j({\bf k})\Psi_{j^{\prime}}({\bf k}^{\prime})\;.
\end{eqnarray}
It follows from Eqs.\ (\ref{sexpansion}-\ref{eliashbergfct})
that the self--energy coefficients $\Sigma_{i,j}^{{\rm R},{\rm A}}(\omega)$
are given by
\begin{eqnarray}
\Sigma_{i,j}^{{\rm R},{\rm A}}(\omega)
&=&
\sum\limits_{j^{\prime}}\,\int\limits_{0}^{\infty}{\rm d}\/\Omega\:
\alpha^2_{j,j^{\prime}}(\Omega)F(\Omega)
L(\omega\pm {\rm i}\/\delta,\Omega)
\;,
\end{eqnarray}
where
\begin{eqnarray}
\alpha^2_{j,j^{\prime}}(\Omega)F(\Omega)
&=&
\frac{1}{N(0)}\,
\sum\limits_{{\bf k}}
\sum\limits_{{\bf k}^{\prime}}
\delta(\epsilon_{{\bf k}})
\delta(\epsilon_{{\bf k}^{\prime}})
\alpha^2({\bf k},{\bf k}^{\prime},\Omega)F(\Omega)
\Psi_j({\bf k})\Psi_{j^{\prime}}({\bf k}^{\prime})
\;.
\end{eqnarray}
Here $N(0)$ denotes the density of electron
states on the Fermi surface.
The first Fermi harmonic is $\Psi_0({\bf k})=1$ and the function
\begin{eqnarray}
\alpha^2_{0,0}(\Omega)F(\Omega)
&=&
\frac{1}{N(0)}\,
\sum\limits_{{\bf k}}
\sum\limits_{{\bf k}^{\prime}}
\delta(\epsilon_{{\bf k}})
\delta(\epsilon_{{\bf k}^{\prime}})
\alpha^2({\bf k},{\bf k}^{\prime},\Omega)F(\Omega)
\end{eqnarray}
is obviously the well known Eliashberg spectral
function which determines the superconductivity of
metals in the simple $s$--pairing case.
Introducing the real and imaginary parts of the self--energy,
$\Sigma_{1}({\bf k},\omega)$ and $\Sigma_{2}({\bf k},\omega)$,
respectively, the Green's function becomes
\begin{eqnarray}\label{greensrealimag}
G^{-1}({\bf k},\omega+{\rm i}\/\delta)
&=&
\omega-\epsilon_{\bf k}-\Sigma_{1}({\bf k},\omega+{\rm i}\/\delta)
-{\rm i}\/\Sigma_{2}({\bf k},\omega+{\rm i}\/\delta)
\;.
\end{eqnarray}
The pole of the Green's function determines the spectrum of
one--particle excitations. At small energies Eq.\ (\ref{greensrealimag})
can be rewritten as
\begin{eqnarray}\label{greensrealimag2}
G^{-1}({\bf k},\omega+{\rm i}\/\delta)
&=&
\omega\left(1-
\left.
\frac{\partial \Sigma_{1}({\bf k},\omega)}{\partial \omega}
\right|_{\omega=0}\right) 
-{\rm i}\/\/\Sigma_{2}({\bf k},\omega+{\rm i}\/\delta)
\;.
\end{eqnarray}
Then the pole of $G$ occurs at $\omega_0$, which is given by
\begin{eqnarray}
\omega_0
&=&
E_{\bf k}-\frac{\rm i}{2\tau_{\bf k}}
\;,\\
E_{\bf k}
&=&
\left(1-\frac{\partial \Sigma_{1}({\bf k},\omega)}{\partial \omega}\right)^{-1}
\epsilon_{\bf k}
\;,\\\label{oneparticlerelaxation}
\frac{1}{2\tau_{\bf k}}
&=&
-\left(1-\frac{\partial \Sigma_{1}({\bf k},\omega)}{\partial 
\omega}\right)^{-1}
\Sigma_{2}({\bf k},E_{\bf k})
\;.
\end{eqnarray}
Here
\begin{eqnarray}
\lambda_{\bf k}
&=&
-\left. \frac{\partial \Sigma_{1}({\bf k},\omega)}{\partial 
\omega}\right|_{\omega=0}
\end{eqnarray}
describes the change of the effective mass of
the electron, while $\frac{1}{2\tau_{\bf k}}$ describes their 
relaxation rate.
All functions can be rewritten in
terms of Fermi harmonics and the spectral function
$\alpha^2_{j,j^{\prime}}(\Omega)F(\Omega)$. We shall turn
to this problem later when we discuss the conductivity
of metals in the presence of electron--phonon interaction.

As usual we write the conductivity of a metal in the presence
of electron--phonon interaction in terms of the analytically
continued electromagnetic kernel $K_{\alpha\beta}(\omega)$,
\begin{eqnarray}\label{kernel}
\sigma_{\alpha\beta}(\omega) &=& 
\frac{e^2K_{\alpha\beta}(\omega)}{4\pi\/{\rm i}\/\omega}\;,
\end{eqnarray}
which, in turn, is expressed through the one--particle 
Green's function $G_{i}({\bf k},\omega)$ and corresponding vertex function
$\Gamma_{\beta}$. In the framework of the thermodynamical
theory of the perturbations the expression for $K_{\alpha\beta}(\omega)$
is
\begin{eqnarray}\label{kernel2}
K_{\alpha\beta}(\/{\rm i}\/\omega_n)
&=&
{\rm T}\sum\limits_{\omega_{n^{\prime}}}
\sum\limits_{{\bf k}^{\prime}}\upsilon^{\alpha}_{{\bf k}^{\prime}}\,
G({\bf k}^{\prime},{\rm i}\/\omega_{n^{\prime}})
G({\bf k}^{\prime},{\rm i}\/\omega_{n^{\prime}}+{\rm i}\/\omega)
\times\\\nonumber &&\times
\Gamma_{\beta}({\bf k}^{\prime},{\rm i}\/\omega_{n^{\prime}},
{\rm i}\/\omega_{n^{\prime}}+{\rm i}\/\omega)\;,
\end{eqnarray}
where $\upsilon^{\alpha}_{{\bf k}^{\prime}}$ is the $\alpha$--component
of the electron velocity. 
Using the ladder diagram approximation the equation for the vertex
function is written as \cite{holstein}
\begin{eqnarray}\label{vortex}
\Gamma_{\beta}({\bf k}^{\prime},{\rm i}\/\omega_{n^{\prime}},
{\rm i}\/\omega_{n^{\prime}}+{\rm i}\/\omega_n)
&=&
\upsilon^{\beta}_{{\bf k}^{\prime}}+
{\rm T}\sum\limits_{{\bf k}^{\prime\prime},n^{\prime\prime}}
\left| \left\langle
g_{{\bf k}^{\prime\prime}}({\bf k}^{\prime} - {\bf k}^{\prime\prime},
i, i^{\prime}, \lambda)
\right\rangle
\right|^2
\times\\\nonumber &&\times
D_{\lambda}({\bf k}^{\prime} - {\bf k}^{\prime\prime},
{\rm i}\/\omega_{n^{\prime}}-{\rm i}\/\omega_{n^{\prime\prime}})
G({\bf k}^{\prime\prime},{\rm i}\/\omega_{n^{\prime\prime}})
G({\bf k}^{\prime\prime},{\rm i}\/\omega_{n^{\prime\prime}}
+{\rm i}\/\omega_n)
\times\\\nonumber &&\times
\Gamma_{\beta}({\bf k}^{\prime\prime},{\rm i}\/\omega_{n^{\prime\prime}},
{\rm i}\/\omega_{n^{\prime\prime}}+{\rm i}\/\omega_n)
\;.
\end{eqnarray}
Before we solve Eqs.\ (\ref{kernel2}) and (\ref{vortex}) we
simplify them somewhat. Firstly, as the conductivity of
any metal in the absence of a magnetic field can be diagonalised
in the appropriate representation we omit in the following
the indices $\alpha$ and $\beta$ considering the conductivity as a 
scalar.
This is done bearing in mind that the
absolute value of the conductivity in 
non--cubic crystals is anisotropic and that the corresponding 
functions determining the temperature and frequency 
dependence of $\sigma(\omega,{\rm T})$ reflect this anisotropy. Secondly, 
we omit the electron band indexes taking into account that interband
transitions can be calculated separately if needed. We also omit
the electron spin index multiplying by two the sum over the electron
momentum ${\bf k}$.
After that Eq.\ (\ref{vortex}) is rewritten in a simplified form
\begin{eqnarray}
\Gamma_x({\bf k}^{\prime},{\rm i}\/\omega_{n^{\prime}},
{\rm i}\/\omega_{n^{\prime}}+{\rm i}\/\omega_n)
&=&
\upsilon^{x}_{{\bf k}^{\prime}}+
2{\rm T}\sum\limits_{{\bf k}^{\prime\prime}}
\sum\limits_{n^{\prime\prime}}
\upsilon^{x}_{{\bf k}^{\prime\prime}}
\int\limits_{-\infty}^{\infty}{\rm d}\/\epsilon\:
\delta(\epsilon - \epsilon_{{\bf k}^{\prime\prime}})
\times\\\nonumber &&\times
\frac{1}{\pi}
\int\limits_{0}^{\infty}{\rm d}\/\Omega\:
\alpha^2({\bf k}^{\prime},{\bf k}^{\prime\prime})F(\Omega)
\times\\\nonumber &&\times
\left(
\frac{1}{{\rm i}\/\omega_{n^{\prime}} - 
{\rm i}\/\omega_{n^{\prime\prime}} - \Omega}-
\frac{1}{{\rm i}\/\omega_{n^{\prime}} - 
{\rm i}\/\omega_{n^{\prime\prime}} + \Omega}
\right)
\times\\\nonumber &&\times
G({\bf k}^{\prime\prime},{\rm i}\/\omega_{n^{\prime\prime}})
G({\bf k}^{\prime\prime},{\rm i}\/\omega_{n^{\prime\prime}}
+{\rm i}\/\omega_n)
\times\\\nonumber &&\times
\Gamma_{x}({\bf k}^{\prime\prime},{\rm i}\/\omega_{n^{\prime\prime}},
{\rm i}\/\omega_{n^{\prime\prime}}+{\rm i}\/\omega_n)
\;.
\end{eqnarray}

To establish some important steps in the derivation of 
the general formula for $\sigma(\omega)$ we consider, as a first step,
the case where the vertex correction to the
bare vertex $\Gamma_{\bf k}^{x}$ can be neglected. Then the 
expression for the electromagnetic response kernel 
$K({\rm i}\/\omega_n)$ becomes
\begin{eqnarray}
K_{xx}({\rm i}\/\omega_n)&=&
2{\rm T}\sum\limits_{{\bf k}^{\prime}}
\sum\limits_{\omega_{n^{\prime}}}
\int\limits_{-\infty}^{\infty}{\rm d}\/\epsilon\:
\delta(\epsilon - \epsilon_{{\bf k}^{\prime\prime}})
(\upsilon_{{\bf k}^{\prime}}^{x})^2
\times\\\nonumber &&\times
G({\bf k}^{\prime},{\rm i}\/\omega_{n^{\prime}})
G({\bf k}^{\prime},{\rm i}\/\omega_{n^{\prime}}
+{\rm i}\/\omega_n)
\;.
\end{eqnarray}
We can now use the Poisson summation formula
\begin{eqnarray}
\sum\limits_{n=-\infty}^{\infty}F({\rm i}\/\omega_n)
&=&
-\frac{1}{2\pi\/{\rm i}\/\/{\rm T}\/}\int\limits_{ C}{\rm d}\/\omega\:
\frac{F(\omega)}{{\rm e}^{\frac{\omega}{\/{\rm T}\/}}+1}\;'
\end{eqnarray}
where the contour ${ C}$ encircles the imaginary 
$\omega$--axis. After that we expand the $\omega$--contour
to infinity, picking up contributions from the singularities
of our integrand at ${\rm i}\/\omega_n=\epsilon_{{\bf k}^{\prime}}$
and ${\rm i}\/\omega_n=\epsilon_{{\bf k}^{\prime}}-{\rm i}\/\omega_n$.
As a result we find after rather lengthy but simple
calculation the analytically continued electromagnetic kernel
as
\begin{eqnarray}
K(\omega)&=&
2\sum\limits_{{\bf k}^{\prime}}(\upsilon_{{\bf k}^{\prime}}^{x})^2
\int\limits_{-\infty}^{\infty}{\rm d}\/\epsilon\:
\delta(\epsilon - \epsilon_{{\bf k}^{\prime\prime}})
\int\limits_{-\infty}^{\infty}{\rm d}\/\omega^{\prime}\:
\times\\\nonumber &&\times
\left(
{\rm tanh}\left(\frac{\omega^{\prime}}{2\/{\rm T}}\right) -
{\rm tanh}\left(\frac{\omega^{\prime}+\omega}{2\/{\rm T}}\right) 
\right)\times
\\\nonumber &&\times
\Pi_{0}^{\rm RA}({\bf k}^{\prime},\omega^{\prime},\omega)\;.
\end{eqnarray}
Here $\Pi_{0}^{\rm RA}({\bf k}^{\prime},\omega^{\prime},\omega)$ has
the form
\begin{eqnarray}
\Pi_{0}^{\rm RA}({\bf k}^{\prime},\omega^{\prime},\omega)&=&
G^{\rm R}({\bf k}^{\prime},\omega^\prime+\omega)
G^{\rm A}({\bf k}^{\prime},\omega^\prime)\;,
\end{eqnarray}
where $G^{\rm R}({\bf k}^{\prime},\omega^\prime+\omega)$ and 
$G^{\rm A}({\bf k}^{\prime},\omega^\prime)$
are the retarded and advanced Green's function, respectively.
Their
self--energy parts are given by
Eqs.\ (\ref{selfenergy} - \ref{l2}).
Just as in the case of the one--particle Green's function the
relevant values of $\omega$ are small in comparison to the 
Fermi energy, the value of $\epsilon$ in $\delta(\epsilon
-\epsilon_{\rm k})$ can be neglected, and the integration over
$\epsilon$ can be carried out.
As result we find for the conductivity
\begin{eqnarray}\label{s}
\sigma(\omega)&=&
\frac{e^2}{4\pi\/{\rm i}\/\omega}
\sum\limits_{{\bf k}}(\upsilon_{{\bf k}}^{x})^2
\delta(\epsilon_{\bf k})
\int\limits_{-\infty}^{\infty}{\rm d}\/\omega^{\prime}\:
\left(
{\rm tanh}\left(\frac{\omega^{\prime}+\omega}{2\/{\rm T}}\right) -
{\rm tanh}\left(\frac{\omega^{\prime}}{2\/{\rm T}}\right) 
\right)\times
\\\nonumber &&\times
\Pi_{0}({\hat{\bf k}}_{\rm F},\omega^{\prime},\omega)\;,
\end{eqnarray}
where the function $\Pi_{0}({\hat{\bf k}}_{\rm F},\omega^{\prime},\omega)$
depends on the position of the momentum ${\hat{\bf k}}_{\rm F}$ on
the Fermi surface and is given by
\begin{eqnarray}\label{p0def}
\Pi_{0}({\hat{\bf k}}_{\rm F},\omega^{\prime},\omega)
&=&
\frac{1}{\omega+\Sigma^{\rm R}({\hat{\bf k}}_{\rm F},\omega+\omega^{\prime})
-\Sigma^{\rm A}({\hat{\bf k}}_{\rm F},\omega^{\prime})}\;.
\end{eqnarray}

For the quasi--isotropic case we get, of course, the well known result
for the optical conductivity \cite{allen2,fuenfzehn}
\begin{eqnarray}\label{s1}
\sigma(\omega)&=&
\frac{\omega_{\rm pl}^{2}}{4\pi\/{\rm i}\/\omega}\,
\int\limits_{-\infty}^{\infty}{\rm d}\/\omega^{\prime}\:
\left(
{\rm tanh}\left(\frac{\omega^{\prime}+\omega}{2\/{\rm T}}\right) -
{\rm tanh}\left(\frac{\omega^{\prime}}{2\/{\rm T}}\right) 
\right)
\Pi_{0}(\omega,\omega^{\prime})\;,
\end{eqnarray}
where the plasma frequency of electrons $\omega_{\rm pl}$ is
\begin{eqnarray}
\omega_{\rm pl}^{2}
&=&
2e^2\sum\limits_{\rm k}(\upsilon_{{\bf k}}^{x})^2\delta(\epsilon_{\bf k})
\end{eqnarray}
and the function $\Pi_{0}(\omega,\omega^{\prime})$ is 
\begin{eqnarray}\label{s3}
\Pi_{0}(\omega,\omega^{\prime})
&=&
\frac{1}{\omega+\Sigma^{\rm R}(\omega+\omega^{\prime})
-\Sigma^{\rm A}(\omega^{\prime})+
\frac{\rm i}{\tau_{\rm imp}}}\;.
\end{eqnarray}
Here we introduced the relaxation rate from
impurity scattering $\frac{1}{\tau_{\rm imp}}$.

For the anisotropic case we should expand all functions under the
integral in Eq.\ (\ref{s}) over the Fermi harmonics. The value
$(\upsilon_{{\bf k}}^{x})^2$ is just the square of the Fermi harmonic 
of the order $N=1$ (for details see Ref.\ \cite{allen2}).
The expansion of the function 
$\Pi_{0}({\hat{\bf k}}_{\rm F},\omega^{\prime},\omega)$
can be written as
\begin{eqnarray}
\Pi_{0}({\hat{\bf k}}_{\rm F},\omega^{\prime},\omega)
&=&
\sum\limits_{j}\Pi_{j}(\omega^{\prime},\omega)F_{j}({\hat{\bf k}}_{\rm F})
\;.
\end{eqnarray}
The non--zero result for the conductivity will arise from the first
harmonic, $F_{j}({\hat{\bf k}}_{\rm F}) =1$, and
from the harmonics with even order
$N\ge 2$. The first harmonic gives the same result as 
found for the isotropic case. An example of a higher harmonic is the one which
transforms as the
representation $\Gamma_{12}$ of the crystal symmetry. It has the form
\begin{eqnarray}
\Psi_{j}^{\Gamma_{12}}
&=&
\frac{v_{yx}^2-v_{xy}^2}{\left\langle(v_x^2-v_y^2)^2\right\rangle^{1/2}}
\;.
\end{eqnarray}
At this point we are not certain to which extent the higher harmonics
are relevant in HTSC systems but it is known that their influence is
small in normal metals.
This question should be considered in more detail
but is beyond the scope of this paper.

While the analytic continuation of Eqs. (\ref{kernel}), (\ref{kernel2})
in the zero--order approximation and absence of the
vertex function $\Gamma({\bf k},{\rm i}\/\omega_n,{\rm i}\/\omega_m)$ 
is straightforward, it becomes a non--trivial task in the presence
of $\Gamma({\bf k},{\rm i}\/\omega_n,{\rm i}\/\omega_m)$.
The difficulty arises from the existence of a manifold of functions
which can be obtained as a result of analytical continuation for
one variable while another is fixed. However, this difficulty can be
avoided by changing, as above, the sum over the Matsubara
frequencies to a contour integral. The contour  consists of (three)
circuits around the imaginary axis of the variable $\omega^{\prime}$,
avoiding all
poles and branch cuts of the integrand. Using the methods developed
in Refs.\ \cite{holstein,sechzehn} we obtain
\begin{eqnarray}
\sigma(\omega)
&=&
\frac{2}{4\pi\/{\rm i}\/\omega}\,\sum\limits_{\bf k}
\int\limits_{-\infty}^{\infty}{\rm d}\/\epsilon\:
\delta(\epsilon-\epsilon_{\bf k})\upsilon_{{\bf k}}^{x}
\int\limits_{-\infty}^{\infty}
\frac{{\rm d}\/\omega^{\prime}}{2\pi\/{\rm i}}\:
\left(
{\rm tanh}\left(\frac{\omega^{\prime}+\omega}{2\/{\rm T}}\right) -
{\rm tanh}\left(\frac{\omega^{\prime}}{2\/{\rm T}}\right) 
\right)
\times\\\nonumber &&\times
\Pi_{0}^{{\rm R}{\rm A}}({\bf k},\omega^{\prime},\omega)
\Gamma_{x}({\bf k},\omega^{\prime},\omega)
\end{eqnarray}
and
\begin{eqnarray}\label{gammaselfconsistent}
\Gamma_{x}({\bf k},\omega^{\prime},\omega)
&=&
\upsilon_{{\bf k}}^{x}+2\,\sum\limits_{{\bf k}^{\prime}}
\int\limits_{-\infty}^{\infty}{\rm d}\/\epsilon\:
\delta(\epsilon-\epsilon_{{\bf k}^{\prime}})
\int\limits_{-\infty}^{\infty}
\frac{{\rm d}\/\omega^{\prime\prime}}{2\pi\/{\rm i}}\:
\times\\\nonumber &&\times
\left[
{\rm tanh}\left(\frac{\omega^{\prime\prime}+\omega}{2\/{\rm T}}\right)
\lambda_{{\bf k}{\bf k}^{\prime}}(\omega^{\prime}-
\omega^{\prime\prime} +{\rm i}\/\delta)
-
{\rm tanh}\left(\frac{\omega^{\prime\prime}}{2\/{\rm T}}\right)
\lambda_{{\bf k}{\bf k}^{\prime}}(\omega^{\prime\prime}-
\omega^{\prime} -{\rm i}\/\delta)
+\right.
\\\nonumber&&
\left. +
{\rm coth}\left(\frac{\omega^{\prime}-\omega^{\prime\prime}}{
2\/{\rm T}}\right)\left(
\lambda_{{\bf k}{\bf k}^{\prime}}(\omega^{\prime\prime}-
\omega^{\prime} +{\rm i}\/\delta) - 
\lambda_{{\bf k}{\bf k}^{\prime}}(\omega^{\prime}-
\omega^{\prime\prime} -{\rm i}\/\delta)
\right)\right]
\times\\\nonumber &&\times
\Pi_{0}^{{\rm R}{\rm A}}({\bf k}^{\prime},\omega^{\prime\prime},\omega)
\Gamma_{x}({\bf k}^{\prime},\omega^{\prime\prime},\omega)
\;,
\end{eqnarray}
where the function
\begin{eqnarray}
\lambda_{{\bf k}{\bf k}^{\prime}}(\omega)
&=&
\int\limits_{0}^{\infty}{\rm d}\/\Omega\:
\alpha^2({\bf k},{\bf k}^{\prime},\Omega)F(\Omega)
\left[
\frac{1}{\omega-\Omega}-\frac{1}{\omega+\Omega}
\right]
\end{eqnarray}
was introduced.
Taking into account that apart from
$\Pi_{0}({\bf k}^{\prime},\omega^{\prime\prime},\omega)$
all functions under the integral on the
right--hand side of Eq.\ (\ref{gammaselfconsistent})
depend only weakly on the variable $\epsilon_{{\bf k}^{\prime}}$ 
we can integrate over this variable and find
\begin{eqnarray}
\Gamma_{x}({\hat{\bf k}}_{\rm F},\omega^{\prime},\omega)
&=&
\upsilon_{{\bf k}}^{x}+\sum\limits_{{\bf k}^{\prime}}
\delta(\epsilon_{{\bf k}^{\prime}})
\int\limits_{-\infty}^{\infty}
{\rm d}\/\omega^{\prime\prime}\:
\Pi_{0}({\hat{\bf k}}^{\prime}_{\rm F},\omega^{\prime\prime},\omega)
\times\\\nonumber &&\times
\left[
I(\omega^{\prime}-{\rm i}\/\delta,\Omega,\omega^{\prime\prime})
-
I(\omega^{\prime}+\omega+{\rm i}\/\delta,\Omega,\omega^{\prime\prime})
\right]
\Gamma_{x}({\hat{\bf k}}^{\prime}_{\rm F},\omega^{\prime\prime},\omega)
\;.
\end{eqnarray}
Here $\Pi_{0}({\hat{\bf k}}^{\prime}_{\rm F},\omega^{\prime\prime},\omega)$
is defined by Eq.\ (\ref{p0def}) and $I(\omega,\Omega,\omega^{\prime})$
is
\begin{eqnarray}
I(\omega,\Omega,\omega^{\prime})
&=&
\frac{1-f(\omega^{\prime})+N(\Omega)}{\omega-\Omega-\omega^{\prime}}
+
\frac{f(\omega^{\prime})+N(\Omega)}{\omega+\Omega-\omega^{\prime}}
\;.
\end{eqnarray}
At this stage it is useful to introduce a new function
$\gamma_x({\hat{\bf k}}_{\rm F},\omega^{\prime},\omega)$,
\begin{eqnarray}
\gamma_x({\hat{\bf k}}_{\rm F},\omega^{\prime},\omega)
&=&
\Pi_{0}({\hat{\bf k}}_{\rm F},\omega^{\prime},\omega)
\Gamma_{x}({\hat{\bf k}}_{\rm F},\omega^{\prime},\omega)
\;.
\end{eqnarray}
In terms of this function the conductivity can be expressed as
\begin{eqnarray}
\sigma(\omega)
&=&
\frac{2}{4\pi\/{\rm i}\/\omega}\,\sum\limits_{\bf k}
\delta(\epsilon_{\bf k})\upsilon_{{\bf k}}^{x}
\int\limits_{-\infty}^{\infty}
{\rm d}\/\omega^{\prime}\:
\left(
{\rm tanh}\left(\frac{\omega^{\prime}+\omega}{2\/{\rm T}}\right) -
{\rm tanh}\left(\frac{\omega^{\prime}}{2\/{\rm T}}\right) 
\right)
\times\\\nonumber &&\times
\gamma_x({\hat{\bf k}}_{\rm F},\omega^{\prime},\omega)
\;.
\end{eqnarray}
For $\gamma_x({\hat{\bf k}}_{\rm F},\omega^{\prime},\omega)$
one can write the equation
\begin{eqnarray}\label{smallgamma}
\omega\gamma_x({\hat{\bf k}}_{\rm F},\omega^{\prime},\omega)
&=&
\upsilon_{{\bf k}}^{x}+\sum\limits_{{\bf k}^{\prime}}
\delta(\epsilon_{{\bf k}^{\prime}})
\int\limits_{0}^{\infty}{\rm d}\/\Omega\: 
\alpha^2({\bf k},{\bf k}^{\prime},\Omega)F(\Omega)
\times\\\nonumber &&\times
\int\limits_{-\infty}^{\infty}{\rm d}\/\omega^{\prime\prime}
\left[
I(\omega^{\prime}-{\rm i}\/\delta,\Omega,\omega^{\prime\prime})-
I(\omega^{\prime}+\omega+{\rm i}\/\delta,\Omega,\omega^{\prime\prime})
\right]
\times\\\nonumber &&\times
\left(
\gamma_x({\hat{\bf k}}_{\rm F}^{\prime},\omega^{\prime\prime},\omega)
-
\gamma_x({\hat{\bf k}}_{\rm F},\omega^{\prime},\omega)\right)
\;.
\end{eqnarray}
Now we use the expansion of the functions in 
Eq.\ (\ref{smallgamma}) in Fermi harmonics. This yields
\begin{eqnarray}\label{smallgamma2}
\omega\gamma_j(\omega^{\prime},\omega)
&=&
\left\langle\upsilon_{x}^2\right\rangle^{1/2}\delta_{jx}+
\sum\limits_{j^{\prime}}
\int\limits_{0}^{\infty}{\rm d}\/\Omega\: 
\times\\\nonumber &&\times
\int\limits_{-\infty}^{\infty}{\rm d}\/\omega^{\prime\prime}
\left[
I(\omega^{\prime}-{\rm i}\/\delta,\Omega,\omega^{\prime\prime})-
I(\omega^{\prime}+\omega+{\rm i}\/\delta,\Omega,\omega^{\prime\prime})
\right]
\times\\\nonumber &&\times
\left(
\alpha^2_{jj^{\prime}}(\Omega)F(\Omega)
\gamma_{j^{\prime}}(\omega^{\prime\prime},\omega)
-\sum\limits_{j^{\prime\prime}} C_{jj^{\prime}j^{\prime\prime}}
\alpha^2_{j^{\prime\prime}0}(\Omega)F(\Omega)
\gamma_{j^{\prime}}(\omega^{\prime\prime},\omega)
\right)
\;,
\end{eqnarray}
where we used the Clebsh--Gordon coefficients
\begin{eqnarray}
C_{jj^{\prime}j^{\prime\prime}}
&=&
\frac{1}{N(0)}\sum\limits_{\bf k}\delta(\epsilon_{{\bf k}})
\Psi_j({\bf k})\Psi_{j^{\prime}}({\bf k})
\Psi_{j^{\prime\prime}}({\bf k})
\;.
\end{eqnarray}
All following calculations are greatly simplified if 
\begin{eqnarray}
\alpha^2_{jj^{\prime}}(\Omega)F(\Omega)
&=&
\frac{1}{N(0)}\sum\limits_{\bf k}\sum\limits_{{\bf k}^{\prime}}
\delta(\epsilon_{{\bf k}})\delta(\epsilon_{{\bf k}^{\prime}})
\alpha^2({\bf k},{\bf k}^{\prime},\Omega)F(\Omega)
\Psi_j({\bf k})\Psi_{j^{\prime}}({\bf k}^{\prime})
\end{eqnarray}
has the diagonal form
\begin{eqnarray}
\alpha^2_{jj^{\prime}}(\Omega)F(\Omega)
&=&
\alpha^2_{j}(\Omega)F(\Omega)\delta_{jj^{\prime}}
\;.
\end{eqnarray}
In ordinary metals this assumption is well satisfied, where 
$\alpha^2({\bf k},{\bf k}^{\prime},\Omega)F(\Omega)$ depends
mainly on the difference of the momenta ${\bf k}-{\bf k}^{\prime}$.
There are also some restrictions on the non--diagonal
matrix elements $\alpha^2_{jj^{\prime}}(\Omega)F(\Omega)$, 
$j \not= j^{\prime}$, connected with the crystal symmetries
(see e.\,g.\ Ref.\ \cite{allen2}). Nevertheless, it is difficult
to say something definite about the accuracy of this assumption 
in HTSC systems without concrete calculations of the functions
$\alpha^2({\bf k},{\bf k}^{\prime},\Omega)F(\Omega)$. 
Such calculations where never made until now. 
We use this approximation as a first step. 
In addition, we restrict ourself to the use of
the first two Fermi harmonics.

It is useful to search for the solution of Eq.\ (\ref{smallgamma2})
in the form
\begin{eqnarray}
\gamma_1(\omega^{\prime}\,\omega)
&=&
\frac{1}{\omega+\Sigma_{\rm tr}^{\rm R}(\omega^{\prime}+\omega,
\omega^{\prime})-
\Sigma_{\rm tr}^{\rm A}(\omega^{\prime},\omega)}
\;,
\end{eqnarray}
where, after lengthy but simple calculations, 
the equations for the transport ``self--energies'' can be written in the
form
\begin{eqnarray}\label{srtr}
\Sigma_{\rm tr}^{\rm R}(\omega^{\prime}+\omega,\omega^{\prime})
&=&
\frac{1}{N(0)}
\sum\limits_{\bf k}\sum\limits_{{\bf k}^{\prime},\lambda}
\int\limits_{0}^{\infty}{\rm d}\/\Omega\:
\int\limits_{-\infty}^{\infty}{\rm d}\/\omega^{\prime\prime}\:
I(\omega^{\prime}+\omega+{\rm i}\/\delta,\Omega,\omega^{\prime\prime})
\left|g_{{\bf k}}({\bf k}^{\prime}-{\bf k},\lambda)\right|^2
\times\\\nonumber &&\times
{\rm Im}\/D_{\lambda}({\bf k}-{\bf k}^{\prime},\omega^{\prime})
\frac{\upsilon_{x}^2}{\left\langle\upsilon_{x}^2\right\rangle}
\left(
1-\frac{\upsilon_x^{\prime}}{\upsilon_{x}}
\frac{\omega+\Sigma_{\rm tr}^{\rm R}(\omega^{\prime}+\omega,
\omega^{\prime})-
\Sigma_{\rm tr}^{\rm A}(\omega^{\prime},\omega)}{\omega+
\Sigma_{\rm tr}^{\rm R}(\omega^{\prime\prime}+\omega,
\omega^{\prime})-
\Sigma_{\rm tr}^{\rm A}(\omega^{\prime\prime},\omega^{\prime})}
\right)
\end{eqnarray}
and
\begin{eqnarray}\label{satr}
\Sigma_{\rm tr}^{\rm A}(\omega^{\prime},\omega)
&=&
\frac{1}{N(0)}
\sum\limits_{\bf k}\sum\limits_{{\bf k}^{\prime},\lambda}
\int\limits_{0}^{\infty}{\rm d}\/\Omega\:
\int\limits_{-\infty}^{\infty}{\rm d}\/\omega^{\prime\prime}\:
I(\omega^{\prime}-{\rm i}\/\delta,\Omega,\omega^{\prime\prime})
\left|g^2_{{\bf k}}({\bf k}^{\prime}-{\bf k},\lambda)\right|
\times\\\nonumber &&\times
{\rm Im}\/D_{\lambda}({\bf k}-{\bf k}^{\prime},\omega^{\prime})
\frac{\upsilon_{x}^2}{\left\langle\upsilon_{x}^2\right\rangle}
\left(
1-\frac{\upsilon_x^{\prime}}{\upsilon_{x}}
\frac{\omega+\Sigma_{\rm tr}^{\rm R}(\omega^{\prime}+\omega,
\omega^{\prime})-
\Sigma_{\rm tr}^{\rm A}(\omega^{\prime},\omega)}{\omega+
\Sigma_{\rm tr}^{\rm R}(\omega^{\prime\prime}+\omega,
\omega^{\prime})-
\Sigma_{\rm tr}^{\rm A}(\omega^{\prime\prime},\omega^{\prime})}
\right)
\;.
\end{eqnarray}
The Eqs.\ (\ref{srtr}) and (\ref{satr}) are still implicitly
integral equations for $\Sigma_{\rm tr}^{{\rm R},{\rm A}}$.
However, the integrand on the right hand side of Eqs.\ 
(\ref{srtr})and (\ref{satr}) depends only weakly on the
functions $\Sigma_{\rm tr}^{{\rm R},{\rm A}}$. The detailed numerical 
analysis of these equations, given by Allen \cite{allen} for the
case ${\rm T} = 0$, has shown that the assumption
\begin{eqnarray}
\frac{\omega-\Sigma_{\rm tr}^{\rm R}(\omega^{\prime}+\omega,
\omega^{\prime})-
\Sigma_{\rm tr}^{\rm A}(\omega^{\prime},\omega)}{\omega-
\Sigma_{\rm tr}^{\rm R}(\omega^{\prime\prime}+\omega,
\omega^{\prime})-
\Sigma_{\rm tr}^{\rm A}(\omega^{\prime\prime},\omega^{\prime})}
&=&1
\end{eqnarray}
is satisfied with good accuracy. 
The only difference in Eq.\ (\ref{srtr}) compared to
Eqs.\ (\ref{s1}) and (\ref{s3}) is the appearance of the
transport ``self--energies'' $\Sigma_{\rm tr}^{{\rm R},{\rm A}}$
instead of the one--particle self--energies.
Accordingly, the equations for $\Sigma_{\rm tr}^{{\rm R},{\rm A}}$ can be
written in a form which largely resembles the equations for the
one--particle self--energies, namely
\begin{eqnarray}\label{str1}
\Sigma_{\rm tr}^{\rm R}(\omega^{\prime}+\omega,\omega)
&=&
\int\limits_{0}^{\infty}{\rm d}\/\Omega\:
\alpha^2_{\rm tr}(\Omega)F(\Omega)
\int\limits_{-\infty}^{\infty}{\rm d}\/\omega^{\prime\prime}\:
I(\omega^{\prime}+\omega+{\rm i}\/\delta,\Omega,\omega^{\prime\prime})
\;,
\\\label{str2}
\Sigma_{\rm tr}^{\rm A}(\omega^{\prime},\omega)
&=&
\int\limits_{0}^{\infty}{\rm d}\/\Omega\:
\alpha^2_{\rm tr}(\Omega)F(\Omega)
\int\limits_{-\infty}^{\infty}{\rm d}\/\omega^{\prime\prime}\:
I(\omega^{\prime}-{\rm i}\/\delta,\Omega,\omega^{\prime\prime})
\;,
\end{eqnarray}
and
\begin{eqnarray}
\alpha^2_{\rm tr}(\Omega)F(\Omega)
&=&
\frac{1}{N(0)}
\sum\limits_{\bf k}\sum\limits_{{\bf k}^{\prime},\lambda}
\delta(\epsilon_{{\bf k}})\delta(\epsilon_{{\bf k}^{\prime}})
\int\limits_{-\infty}^{\infty}{\rm d}\/\omega^{\prime\prime}\:
\left|g^2_{{\bf k}}({\bf k}-{\bf k}^{\prime},\lambda)\right|^2
{\rm Im}\/D_{\lambda}(\Omega)
\times\\\nonumber &&\times
\frac{\upsilon_{{\bf k}}^2}{\left\langle\upsilon_{{\bf k}}^2\right\rangle}
\left(
1-\frac{\upsilon_{{\bf k}^{\prime}}}{\upsilon_{{\bf k}}}
\right)
\;.
\end{eqnarray}
The expression for the conductivity can then be written in a form
which is formally identical to Eq.\ (\ref{s1}) 
obtained without vertex corrections
\begin{eqnarray}\label{sv1}
\sigma(\omega)&=&
\frac{\omega_{\rm pl}^{2}}{4\pi\/{\rm i}\/\omega}\,
\int\limits_{-\infty}^{\infty}{\rm d}\/\omega^{\prime}\:
\left(
{\rm tanh}\left(\frac{\omega^{\prime}+\omega}{2\/{\rm T}}\right) -
{\rm tanh}\left(\frac{\omega^{\prime}}{2\/{\rm T}}\right) 
\right)
\times\\\nonumber &&\times
\frac{1}{\omega+\Sigma_{\rm tr}^{\rm R}(\omega^{\prime}+\omega)-
\Sigma_{\rm tr}^{\rm A}(\omega^{\prime}+\omega)+
\frac{{\rm i}}{\tau_{\rm imp}}}
\;.
\end{eqnarray}
Equation (\ref{sv1}) has just the form which is
normally used for the calculation of the transport properties
of metals (see. e.\,g.\ \cite{zhenya,allen2}).

The expression (\ref{sv1}) for the conductivity can be simplified
even more in the case of weak electron--phonon interaction
\cite{oleg}, to yield the so--called ``extended''
Drude formula
\begin{eqnarray}\label{extended}
\sigma(\omega)
&=&
\frac{\omega_{\rm pl}^{2}}{4\pi}\,
\frac{1}{{\rm i}\/\omega-W(\omega)-\frac{1}{\tau_{\rm imp}}}
\;,
\end{eqnarray}
where
\begin{eqnarray}
W(\omega)
&=&
{\rm i}\/\omega\left(
1-\frac{m^*_{\rm tr}(\omega)}{m}\right)+\frac{1}{\tau_{\rm tr}(\omega)}
\;,
\\
W(\omega)
&=&
-2\/{\rm i}\/\int\limits_{0}^{\infty}{\rm d}\/\Omega\:
\alpha^2_{\rm tr}(\Omega)F(\Omega)\,
K\!\left(\frac{\omega}{2\pi\/{\rm T}},\frac{\Omega}{2\pi\/{\rm T}}\right)
\;,
\end{eqnarray}
and the function
$K\!\left(\frac{\omega}{2\pi\/{\rm T}},\frac{\Omega}{2\pi\/{\rm T}}\right)$
has the form
\begin{eqnarray}
K\left(x,y\right)
&=&
\frac{\rm i}{y}+\left\{\frac{y-x}{x}\left[
\Psi(1-{\rm i}\/x+{\rm i}\/y)-
\Psi(1+{\rm i}\/y)
\right]\right\} -
\left\{ y \longleftrightarrow -y \right\}
\;.
\end{eqnarray}

Firstly, we shall check the accuracy of the approximation of
the expression (\ref{sv1}) for $\sigma(\omega)$ by the ``extended''
Drude formula, Eq.\ (\ref{extended}). For that purpose we use the
transport spectral function $\alpha^2_{\rm tr}(\omega)F(\omega)$
of the form shown in Fig.\ \ref{fig0}. 
A comparison  between Eqs.\ (\ref{sv1}) and (\ref{extended})
is shown in Fig.\ \ref{fig1} for constants of coupling $\lambda_{\rm tr}=1$, 
where
\begin{eqnarray}\label{alphatr}
\lambda_{\rm tr}&=&
2\,\int\limits_{0}^{\infty}\frac{{\rm d}\/\Omega}{\Omega}\;
\alpha^2_{\rm tr}(\Omega)F(\Omega)
\;.
\end{eqnarray}
There are some differences in the low energy regime, $\omega \lesim 200$ 
cm$^{-1}$, shown in Fig.\ \ref{fig2}, especially at low temperatures. 
One should be careful using
the ``extended'' Drude formula in this case. However, even in
this region the difference is rather small. 
The difference between the conductivity calculated by the formulae
(\ref{sv1}) and (\ref{extended}) continues to be small even for 
a constant of coupling $\lambda \simeq 2$.

The transport relaxation rate
\begin{eqnarray}
\frac{1}{\tau^*_{\rm tr}(\omega)}
&=&
\frac{1}{\tau_{\rm tr}(\omega)}\,
\frac{m}{m^*(\omega)}
\end{eqnarray}
has a universal behaviour for metals with different values of
$\lambda$ and different phonon density of states.
Figure \ref{fig3} shows the frequency dependence of the
relaxation rate $\frac{1}{\tau^*_{\rm tr}(\omega)}$ in 
dimensionless form. Here $\omega_{\rm m}$ is the value of the
maximum phonon frequency (i.\,e.\ the end of the phonon spectrum).
The six different curves plotted in Fig.\ \ref{fig3} are
indistinguishable as function of the dimensionless 
energy $\omega/\omega_{\rm m}$.
The quantity $\frac{1}{\tau^*_{\rm tr}(\omega)}$ is also rather
universal as a function of the dimensionless
temperature ${\rm T}/\pi\omega_{\rm m}$. Figure \ref{fig4}
shows $\frac{1}{\tau^*_{\rm tr}(\omega)}$ for ${\rm T} = 10$ K
and $100$ K.
Firstly, we would like to emphasise the quasi--linear 
$\omega$--dependence of $\frac{1}{\tau^*_{\rm tr}(\omega)}$ over a 
large energy interval, $0.5 \le \omega \le (3-4)\,\omega_{\rm m}$.
The function $\frac{1}{\tau^*_{\rm tr}(\omega)}$ increases with
increasing energy $\omega$ up to very high values, $\omega
\simeq 10\,\omega_{\rm m}$. This contrasts strongly with the
behaviour of the one--particle relaxation rate 
$\frac{1}{\tau(\omega)}$  defined by 
Eq.\ (\ref{oneparticlerelaxation}) and shown in the inset of
Fig.\ \ref{fig4}. It is well known \cite{siebzehn} 
that the latter rapidly increases with energy and becomes constant
for $\omega\ge \omega_{\rm m}$.
This difference in the behaviour of the functions
$\frac{1}{\tau^*_{\rm tr}(\omega)}$ and $\frac{1}{\tau(\omega)}$
was first discussed in Ref.\ \cite{shulga}. 

As mentioned above, there are only a few investigations 
\cite{joyce,bednorz} of the influence of the electron--phonon interaction
on the optical spectra of normal state metals where the frequency dependence of
these effects was observed.
Besides the reasons mentioned above there is another very important reason
for the small number of such investigations. Namely, the absolute values of
the frequency dependence of the discussed effects is determined 
by the value of $\frac{1}{\tau^*_{\rm tr}(\omega)}$ at
$\omega\rightarrow\infty$ \cite{shulga}. This function can be written as
\begin{eqnarray}
\lim\limits_{\omega\rightarrow\infty}\frac{1}{\tau^*_{\rm tr}(\omega)}
&=&
\pi\lambda\left\langle\omega\right\rangle
\;,
\end{eqnarray}
where
\begin{eqnarray}
\lambda\left\langle\omega\right\rangle
&=&
2\,\int\limits_{0}^{\infty}{\rm d}\/\omega\;
\alpha^2_{\rm tr}(\omega)F(\omega)
\;.
\end{eqnarray}
This value can be expressed as
\begin{eqnarray}
\frac{1}{\tau^*_{\rm tr}}
&\simeq &
(1-2)\,\lambda\omega_{\rm m}
\;,
\end{eqnarray}
as can be seen from Figs.\ \ref{fig3} and
\ref{fig4},
and it is very small for ordinary metals.
In lead, for example, one finds $\lambda\left\langle\omega\right\rangle
\approx 100$ cm$^{-1}$.
It is extremely difficult to observe this phenomena in the usual reflection
spectra. Therefore, the observation of Holstein processes in lead was
made \cite{joyce,bednorz} using the light scattering inside
a cavity whose walls hold the sample material. It was averaged over
at least $100$ such reflections to increase the accuracy of the experiment.
It was shown \cite{oleg,achzehn} that a
far--infrared measurement at low temperature, containing phonon--induced
structure, can be inverted to give the spectral function of the 
electron--phonon
interaction $\alpha_{\rm tr}^2(\omega)F(\omega)$.
This function was obtained in Ref.\ \cite{achzehn} for lead in very 
good agreement with experimental data from tunnelling measurements.
We shall come back to discuss this observation in some more detail.

\section{Electron--phonon interaction and optical spectra
of HTSC systems}
A large amount of work has been devoted to the study of optical
spectra of HTSC systems
(see for example the reviews \cite{neunzehn,zwanzig}). Investigations
include normal and superconducting state properties, doping dependence, 
the effect of impurities, etc.\,.
Here, we shall restrict our analysis to optimally doped HTSC in the
normal state. In this case the optical spectra of all HTSC materials show
quite similar behaviour: their reflectivity drops nearly linearly 
with energy from $R\simeq 1$ down to $R\simeq 0.1$ at the plasma edge
$\omega^*_{\rm p}$, where the values of $\omega^*_{\rm p}$ vary for different
HTSC materials, $1$ eV $\le\omega^*_{\rm p}\le1.8$ eV \cite{zwanzig}.
Measurements for different HTSC compounds further coincide in showing large
values for the conductivity in the energy interval
$2$ eV $< \omega < 15$ eV and well developed charge fluctuation spectra
(described by the energy loss function 
$-{\rm Im}\left(\frac{1}{\epsilon(\omega)}\right)$) in the energy interval
$5$ eV $< \omega < 40$ eV \cite{einundzwanzig}.
The high energy part
of spectra above a few eV can be well described in terms of
the usual band structure calculations \cite{zwanzig,zhenya1}.
Moreover, the calculations \cite{zhenya1,dreiundzwanzig}
can also describe quite accurately some low energy interband transitions in
YBa$_2$Cu$_3$O$_7$ \cite{vierundzwanzig}. 

The most unusual part of the optical spectra of HTSC systems is
connected with the strong decrease in reflectivity, almost linear in energy, 
in the region
$0\le\omega\le\omega^*_{\rm p}\simeq 1$ eV \cite{fuenfundzwanzig}. 
This behaviour certainly cannot be explained with the simple Drude approach.
Thomas {\it et al} \cite{sechsundzwanzig} proposed two different
ways of analysing such spectra. First, a one--component 
Fermi--liquid approach, using the ``extended'' Drude formula with frequency
dependent mass $m^*(\omega)$ and relaxation rate $\frac{1}{\tau(\omega)}$.
This model successfully describes the heavy fermion systems 
\cite{siebenundzwanzig}. Secondly, a two--component approach, where
the spectrum is decomposed in two components, 
the Drude part and a mid--infrared (MIR) absorption band.
In the latter case there is no unique way of separating the two components. 

As we have discussed above, including the electron--phonon interaction
in the consideration leads immediately to the representation of the
optical conductivity in terms of the ``extended'' Drude formula.  
It has been shown earlier for some HTSC \cite{shulga,holger}
that the existence of strong electron--phonon coupling, 
including some amount of MIR excitations, can indeed
explain their optical conductivity. We now consider this approach
in some more detail. To describe the optical spectra of HTSC systems 
we use the formulae obtained in section \ref{formulae} using
a transport spectral function $\alpha^2_{\rm tr}(\omega)F(\omega)$
of the form shown in Fig.\ \ref{fig0}, multiplied by $\omega^2$.
This spectral function extends up to $\omega_{\rm m} = 735$ cm$^{-1}$ and
well resembles
the phonon spectra of HTSC systems. Generally, the optical properties 
do not depend on the actual shape of $\alpha^2_{\rm tr}(\omega)F(\omega)$
but on some moments of it. Thus, we use the constant of
electron--phonon coupling $\lambda_{\rm tr}$ as defined in Eq.\ 
(\ref{alphatr}) as fit parameter.

Figure \ref{fig6} shows the reflectivity of optimally doped 
La$_{2-x}$Sr$_{x}$CuO$_4$ (LSCO) at room temperature. We have used the values
$\omega_{\rm p}=1.8$ eV and $\epsilon_{\infty}=4.6$ obtained by one of us
\cite{achtundzwanzig} from band structure calculations. The contribution of
the electron--phonon coupling was supposed to be $\lambda_{\rm tr}=2.5$.
The resulting reflectivity corresponds well to measurements 
on good quality films \cite{fuenfundzwanzig}. The conductivity
calculated in Ref.\ \cite{achtundzwanzig} describes well the
optical spectra of LSCO at energies above $\approx 3$ eV, as was
confirmed by experiment \cite{neunundzwanzig}.
Taking all this into account, 
we can conclude that the band structure approach joined 
with strong electron--phonon interaction can explain the overall
behaviour of the optical spectrum of LSCO in the energy range
$0<\omega\le 40$ eV.
Moreover, this approach also explains the temperature dependence of 
the optical spectrum of LSCO quite well. Figure \ref{fig7} 
shows the reflectivity of LSCO in the FIR at temperatures
${\rm T}=100,\; 200,\; 300$ K. The agreement between experimental 
\cite{dreisig} and calculated curves is good considering 
that the only fit parameter used was the constant of coupling
$\lambda_{\rm tr}$. In Fig.\ \ref{fig8} we show the reflectivity
for LSCO up to $\omega\approx 1$ eV for the same temperatures.
The agreement between calculated and measured curves is still
quite good on this larger energy scale. Nevertheless,
we should point out a small discrepancy in the MIR region. 
While this discrepancy is very small at ${\rm T}=100$ K, it
becomes more pronounced with increasing temperature.
Introducing a MIR band, this
can be seen as further evidence for the temperature dependence 
of such a band, as discussed in Ref.\ \cite{holger}.

We now focus on the temperature dependence of the 
optical reflectivity in YBa$_2$Cu$_3$O$_{7-\delta}$ (YBCO), using the
values $\omega_{\rm p}=3$ eV and $\epsilon_{\infty}=5.8$
obtained from band structure calculations \cite{zhenya1}.
Figure \ref{fig9} shows the reflectivity of YBCO from Ref.\ 
\cite{einundreisig} compared to our calculations for 
${\rm T}=100,\; 200,\; 300$ K, where we have used an impurity scattering
rate $\frac{1}{\tau_{\rm imp}}=300$ cm$^{-1}$ in our calculation of the 
relaxation rate. The agreement between experimental and theoretical curves 
seems to be again quite good. The only fit parameter used here
was $\lambda_{\rm tr}=3$. Certainly, there are some discrepancies
between data and theoretical curves in Fig.\ \ref{fig9}. We should
point out that discrepancies of nearly the same order exist between
different type of monocrystals and films. Also, our approach is based on the 
quasi--isotropic approximation for the electron--phonon interaction
which is clearly an oversimplification in these quasi--two dimensional systems.
The inset of Fig.\ \ref{fig9} shows the calculated reflectivity of YBCO up to 
$1.8$ eV at room temperature. It can be seen that there is an upturn
in the reflectivity at MIR frequencies which is certainly much larger
than in LSCO. This confirms the observation made in Ref.\ 
\cite{zweiunddreisig} that the MIR band is more pronounced in YBCO
than in LSCO.
Furthermore, we know from band structure calculations \cite{zhenya1}
that YBCO systems show interband transitions with rather strong intensity
in the MIR region. 

Figure \ref{fig10} shows the behaviour of the transport relaxation
rate $\frac{1}{\tau^*_{\rm tr}(\omega)}$ obtained for YBCO at temperatures
${\rm T}=100,\; 200,\; 300$ K. 
It well resembles the usual shape of such curves
in HTSC and might be compared to Fig.\ 2 in Ref. \cite{dreiundreisig},
where the relaxation rate for Bi$_2$Sr$_2$CaCu$_2$O$_8$ was derived
from the conductivity using the ``extended'' Drude formula. 
It was also shown there that the behaviour of the relaxation rate  
$\frac{1}{\tau^*_{\rm tr}(\omega)}$ cannot be explained in the
framework of marginal Fermi liquid theory \cite{vierundreisig}, since
$\frac{1}{\tau^*_{\rm tr}(\omega)}$ saturates at energies 
$\omega>\omega_{c}\simeq 1500$ cm$^{-1}$ whereas according to
marginal Fermi liquid theory the relaxation rate should continue
in a straight line.
This argument was used in Ref.\ \cite{dreiundreisig} as evidence
against the one--component approach to the optical spectra
of HTSC systems.
However, the saturating behaviour for $\frac{1}{\tau^*_{\rm tr}(\omega)}$
can be clearly seen in Fig.\ \ref{fig10}, at nearly the same energy
as it was observed in Bi$_2$Sr$_2$CaCu$_2$O$_8$. 
Figure \ref{fig11} shows the calculated resistivity for YBCO, demonstrating
clearly a linear temperature dependence over a large temperature interval.

To conclude this considerations, we would like to discuss shortly
the possibility of inverting reflectivity data to obtain
the transport spectral function $\alpha_{\rm tr}^2(\omega)F(\omega)$.
This has been done for lead at ${\rm T} = 0$ \cite{achzehn}, where the
the infrared data was obtained by applying a magnetic field to drive the
system into the normal state. There, a spectral function with detailed
structure could be obtained. 
In HTSC systems the situation is more complex. 
The superconducting state exists up to considerably high temperatures
${\rm T}>{\rm T}_{\rm c}$ and cannot be suppressed by magnetic fields 
to allow IR measurements at ${\rm T} = 0$. At finite temperature the
inversion procedure is far more difficult and as a result such calculations
only yield the rough overall form of the transport spectral function.
Nevertheless, the resulting $\alpha_{\rm tr}^2(\omega)F(\omega)$ 
clearly resemble the phonon spectra of these systems \cite{oleg}.

\section{Conclusion}
The results obtained in this work demonstrate clearly, 
that strong electron--phonon interaction, combined
with band structure calculations, describe
the overall behaviour of the optical spectra and the
main part of the transport properties of HTSC in a
straightforward manner. 
However, we do not claim that this simple quasi--isotropic
approach can explain all details of the behaviour of HTSC systems,
even in the normal state, and even at optimal doping. There are a number
of open problems concerning the behaviour of the Hall coefficient, the 
NMR relaxation rate of the copper sites. We would like to point out
the existence of at least two relaxation rates in the HTSC systems,
namely the quasi--particle relaxation rate 
$\frac{1}{\tau(\omega)}$ and the transport $\frac{1}{\tau^*_{\rm tr}(\omega)}$
which are very different over a large energy range. We cannot rule out
that the relaxation rate involved in the Hall current, due to 
strong and possibly anisotropic electron--phonon interaction, 
will be different from the transport relaxation rate in those
systems. This could lead to the observed temperature dependence of the
Hall coefficient.

Last but not least, we should point out that the simple approach presented
here does not work at low temperatures. It cannot properly describe the
anisotropy of the superconducting order parameter, although it
yields the correct order of magnitude for the value of ${\rm T}_{\rm c}$. There
are additional phenomena besides electron--phonon coupling
which become important at low temperatures.
There are a number of different models which combine
a strong electron--phonon interaction with interband
Coulomb interaction, with the existence of a Van Hove singularity
in the electron spectrum, etc., of which a detailed discussion  
is beyond the scope of this work.
 
\section*{Acknowledgement}
The authors would like to thank O.\ Dolgov for many helpful discussions.
They are also grateful to S.\ Shulga for providing his program for this
calculations. E.\ G.\ M.\ would like to thank the Royal Society and
the Department of Earth Sciences, University of Cambridge, for their
support and kind hospitality during his visit to Cambridge. He
also acknowledges the RFBI for the financial support during the early
stages of this work.

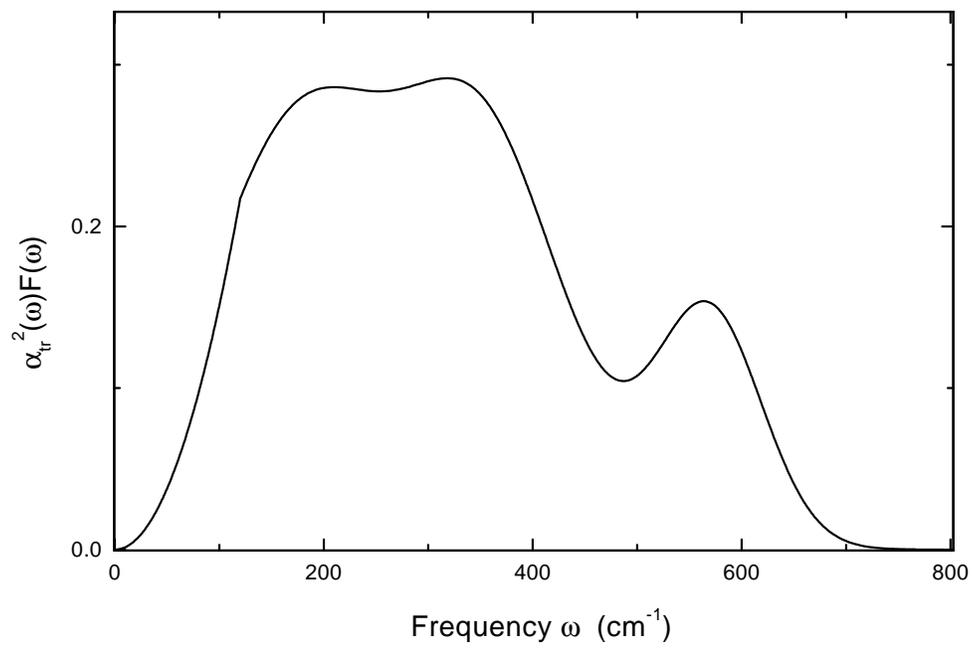
\begin{figure}[ht]
\caption{Transport spectral function $\alpha^2_{\rm tr}(\omega)F(\omega)$
used in the calculations.}
\label{fig0}
\end{figure}

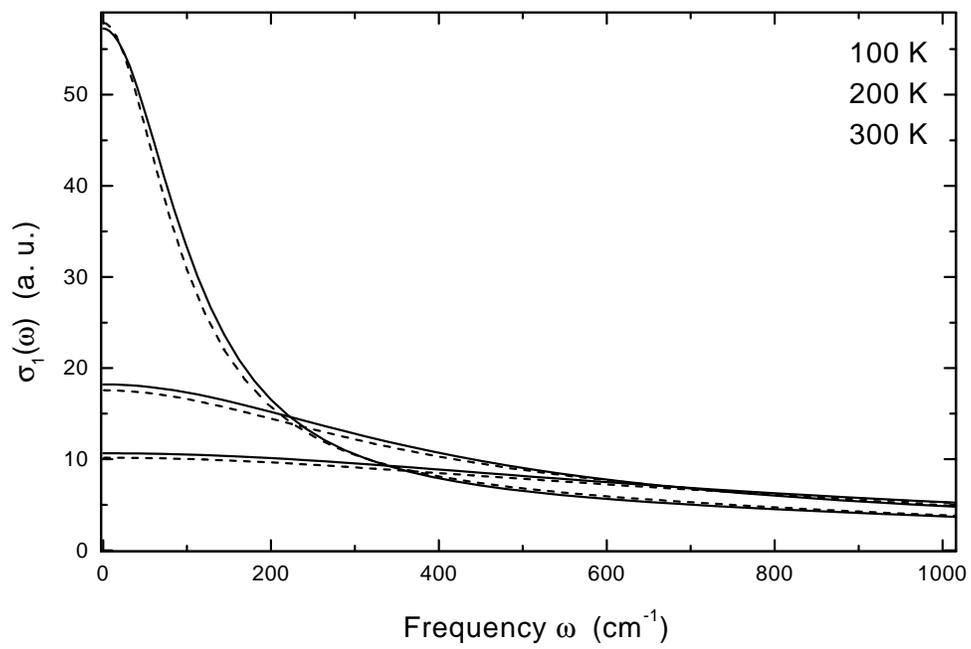
\begin{figure}[ht]
\caption{Comparison between the calculated optical conductivity (real part)
$\sigma_1(\omega)$ using Eq.\ (\ref{sv1}) (dashed line) and 
using the ``extended'' Drude formula, Eq.\ (\ref{extended}) (solid line).}
\label{fig1}  
\end{figure}  

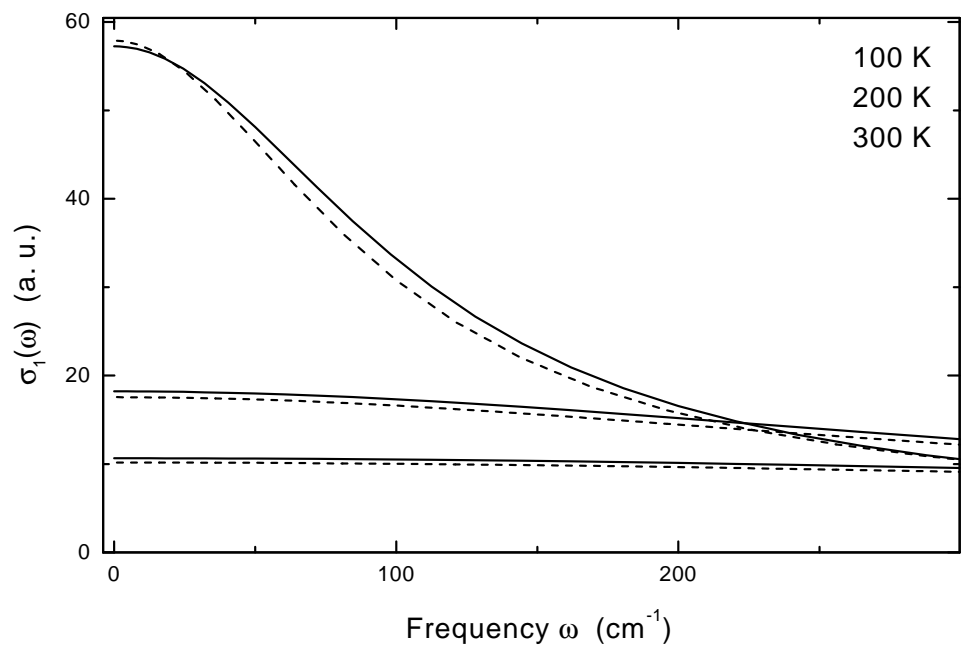
\begin{figure}[ht]
\caption{Same as Fig.\ \ref{fig1} at low energies}
\label{fig2}  
\end{figure} 

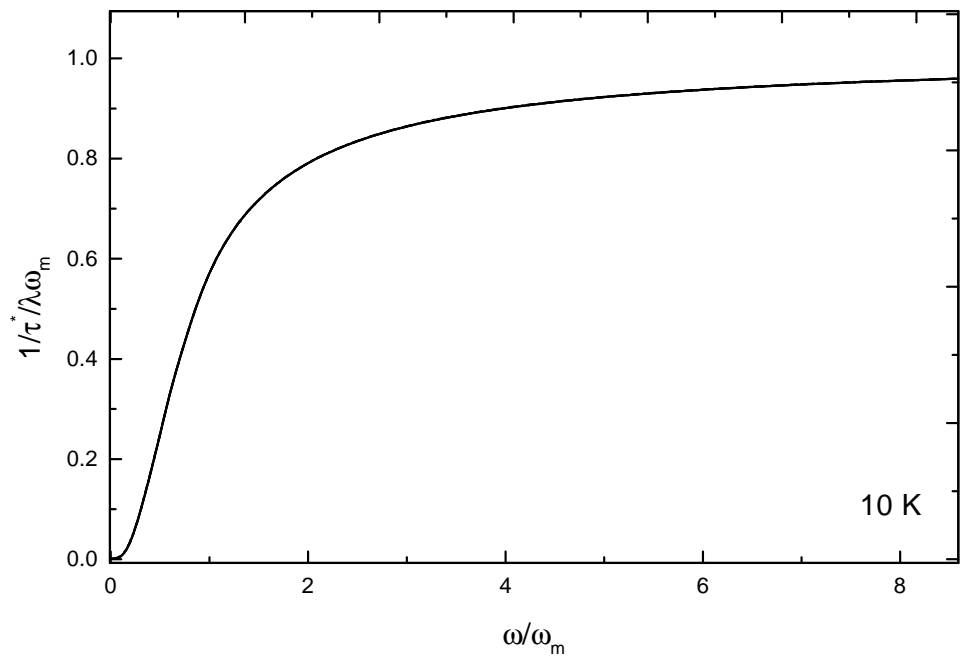
\begin{figure}[ht]
\caption{Transport relaxation rate $\frac{1}{\tau^*_{\rm tr}(\omega)}$ 
at ${\rm T} = 10$ K for
different constants of coupling, $\lambda_{\rm tr} =  
0.2,\; 0.6,\; 1.0,\; 1.4$, and different phonon spectra with
$\omega_{\rm m} = 612$ cm$^{-1}$, $735$ cm$^{-1}$, $882$ cm$^{-1}$,
calculated using the ``extended'' Drude formula. In dimensionless
units all curves fall on top of each other.}
\label{fig3}  
\end{figure} 

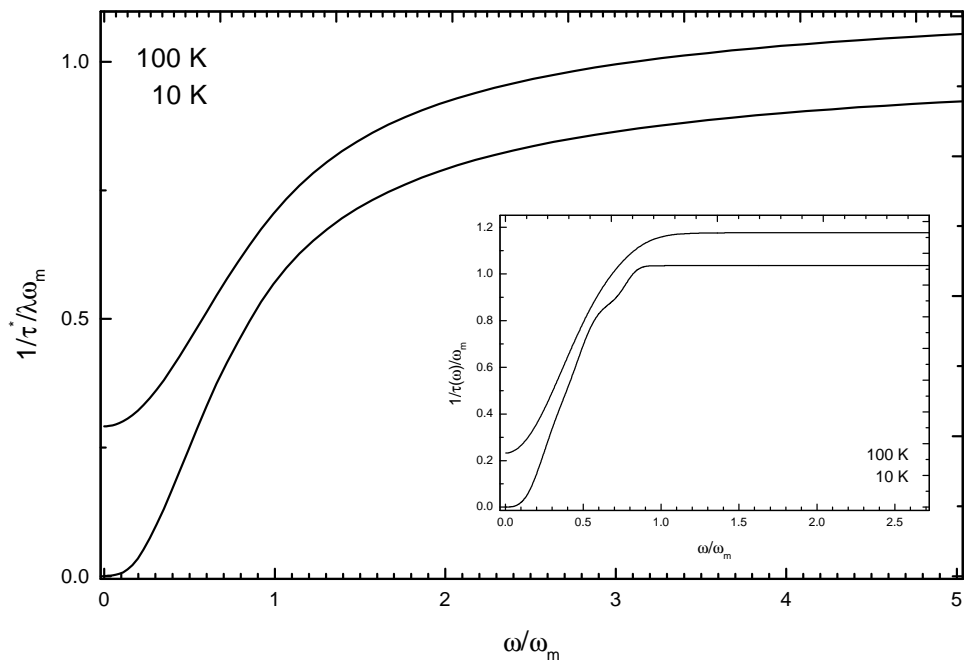
\begin{figure}[ht]
\caption{Transport relaxation rate $\frac{1}{\tau^*_{\rm tr}(\omega)}$
at ${\rm T} = 10$ K and $100$ K for $\lambda =1$, $\omega_{\rm m} =735$ cm$^{-1}$.
The inset shows the one--particle relaxation rate 
calculated for the same parameters.}
\label{fig4}  
\end{figure} 

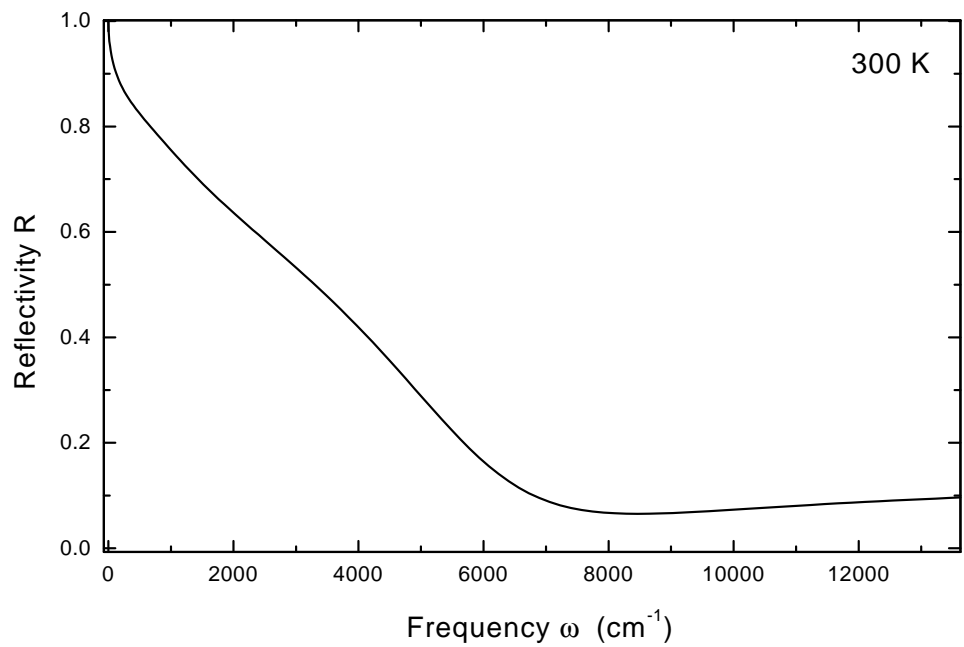
\begin{figure}[ht]
\caption{Calculated reflectivity for optimally doped LSCO at
${\rm T} = 300$ K, using the ``extended'' Drude formula with
$\lambda = 2.5$, $\omega_{\rm p}=1.8$ eV, $\epsilon_{\infty}=4.6$,
$1/\tau_{\rm imp} = 100$ cm$^{-1}$.}
\label{fig6}  
\end{figure} 

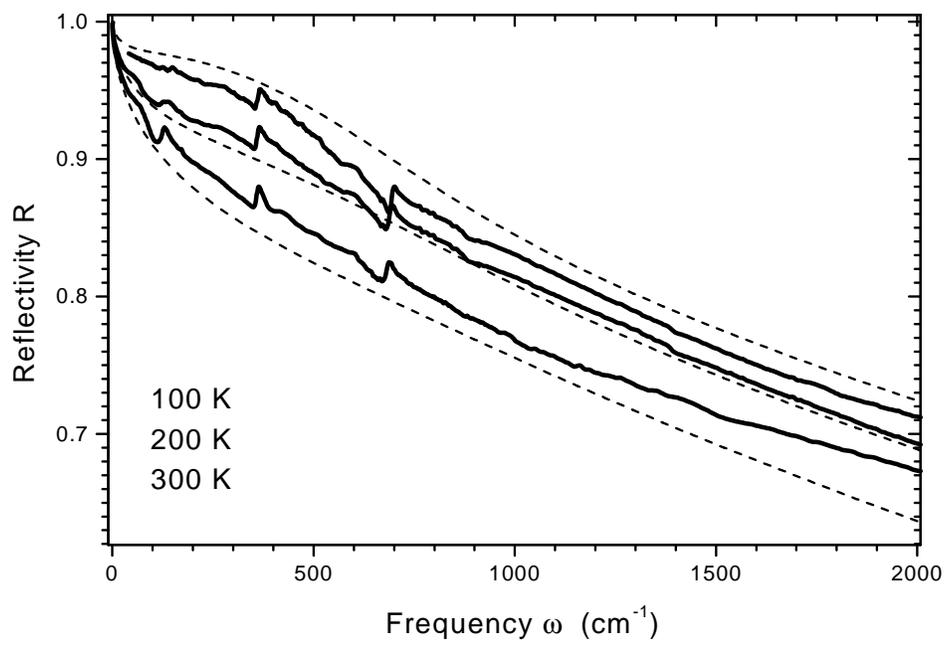
\begin{figure}[ht]
\caption{FIR reflectivity for LSCO from Gao {\it et al} at different
temperatures compared to our calculations.}
\label{fig7}  
\end{figure} 

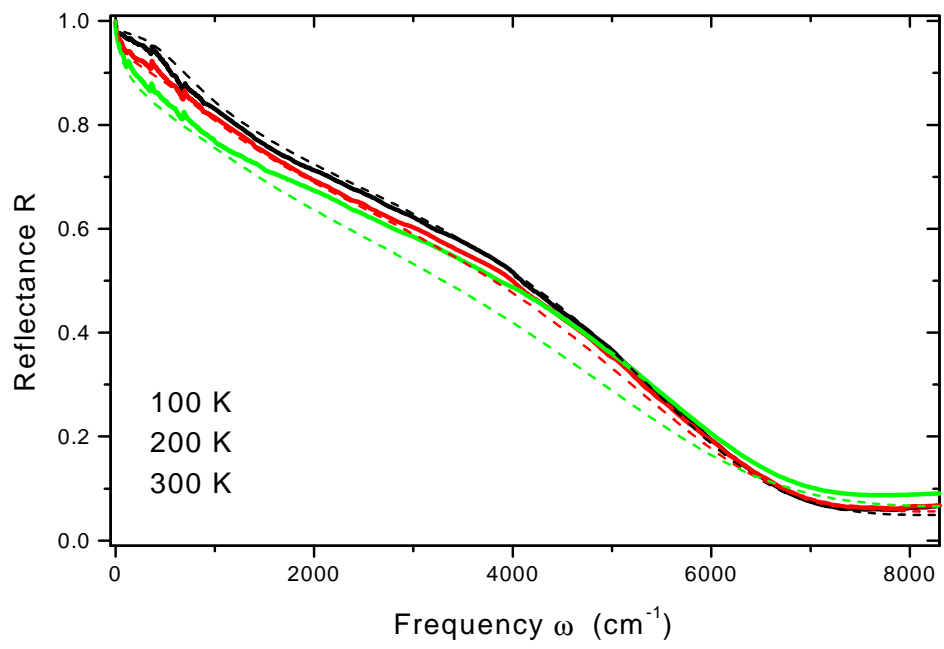
\begin{figure}[ht]
\caption{Same as Fig.\ \ref{fig7} on a larger energy scale.}
\label{fig8}  
\end{figure}

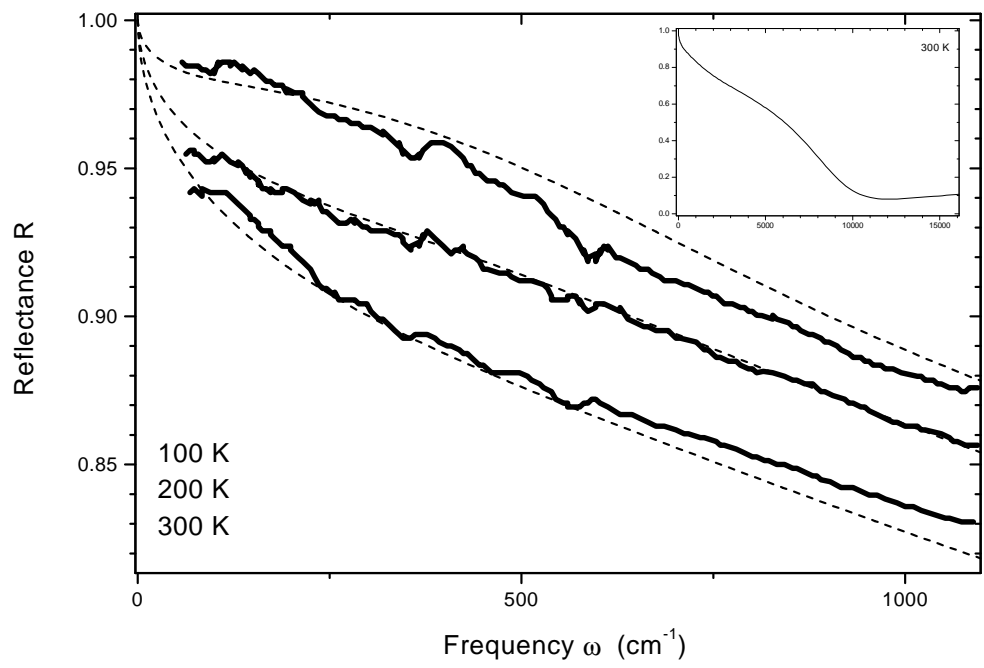
\begin{figure}[ht]
\caption{FIR reflectivity for YBCO from Sch\"utzmann {\it et al} at different
temperatures compared to our calculations, using
$\lambda = 3$, $\omega_{\rm p}=3$ eV, $\epsilon_{\infty}=5.8$,
$1/\tau_{\rm imp} = 300$ cm$^{-1}$. The inset shows the 
calculated curve at room temperature in a larger energy scale.} 
\label{fig9}  
\end{figure}

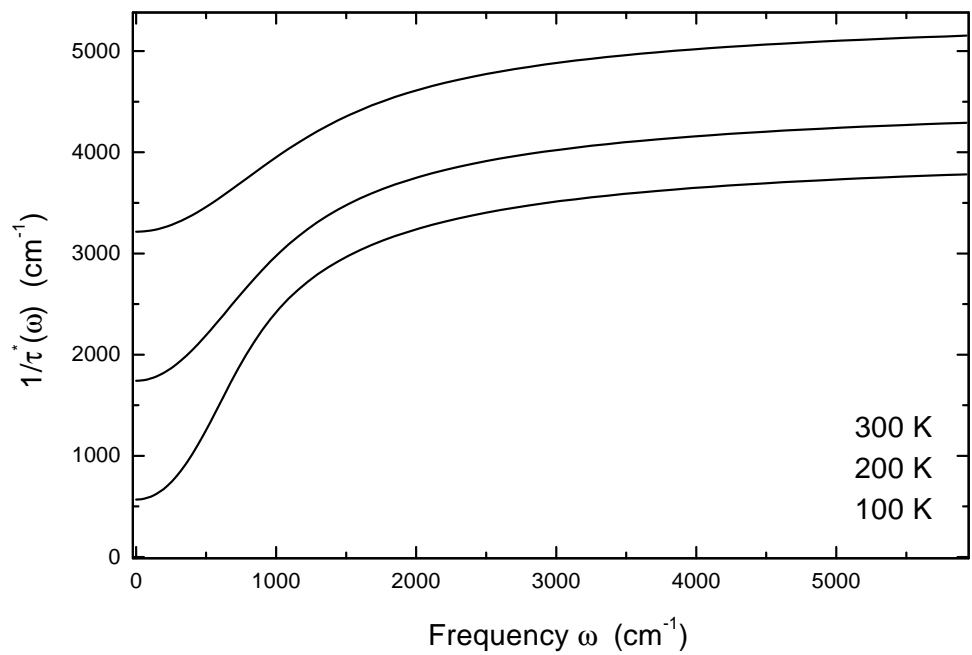
\begin{figure}[ht]
\caption{Calculated transport relaxation rate 
$\frac{1}{\tau^*_{\rm tr}(\omega)}$ for YBCO at different temperatures.}
\label{fig10}  
\end{figure}

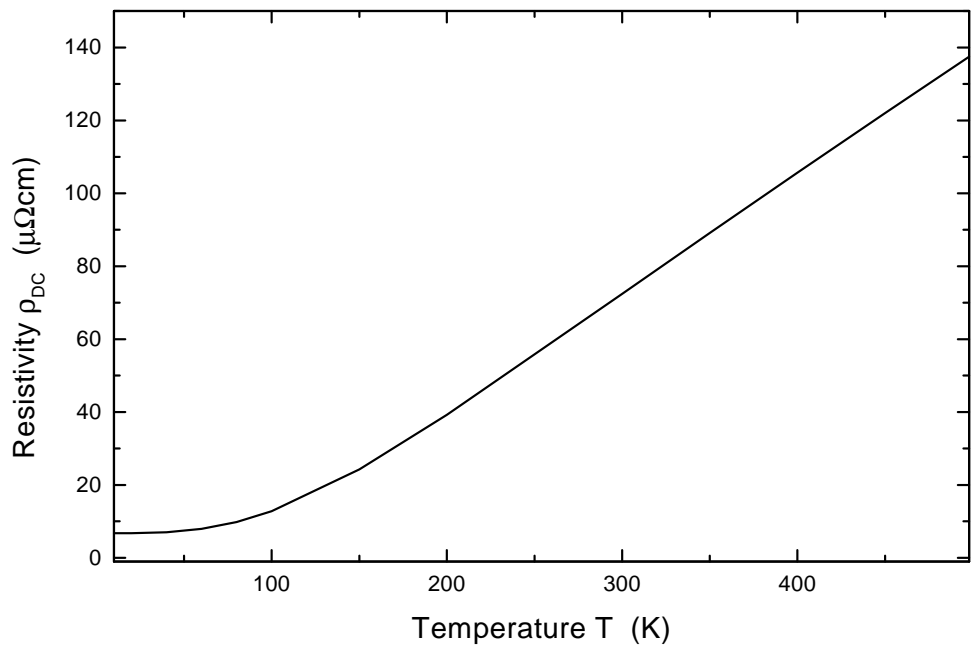
\begin{figure}[ht]
\caption{Calculated DC resistivity for YBCO as a function of temperature.} 
\label{fig11}  
\end{figure}
\end{tighten}
\end{document}